\documentclass[letter]{aa}  

\usepackage{graphicx}
\usepackage{txfonts}
\usepackage{lipsum}
\usepackage{subcaption}         
\usepackage{lscape}             
\usepackage{placeins}           
                                
\usepackage[colorlinks=true,citecolor=blue]{hyperref}
\usepackage{natbib}
\usepackage{orcidlink}
\usepackage{threeparttablex}
\usepackage{CJK}
\providecommand{\linenumbers}{}
\providecommand{\nolinenumbers}{}

\makeatletter
\renewcommand*\aa@textidlineempty{Lei et al.}
\makeatother

\begin{document}
\begin{CJK*}{UTF8}{gbsn}

   \title{Black holes in red-sequence galaxies at $z \sim 0.7-2.5$: Not a source of dark energy, but remnants of little red dots}


%
%
%

   \author{Lei Lei (雷磊)\orcidlink{0000-0003-4631-1915}\inst{1,2}\email{leilei@pmo.ac.cn}
        \and Ze-Fan Wang (王泽凡)\orcidlink{0009-0004-9366-1947}\inst{1,2}\email{zfwang@pmo.ac.cn}
        \and Yi-Ying Wang (王艺颖)\orcidlink{0000-0003-1215-6443}\inst{1}\email{wangyy@pmo.ac.cn}
        \and Lei Feng (冯磊)\orcidlink{0000-0003-2963-5336}\inst{1,2}\corrauth{fenglei@pmo.ac.cn}        
        \and Yi-Zhong Fan (范一中)\orcidlink{0000-0002-8966-6911}\inst{1,2}\corrauth{yzfan@pmo.ac.cn}        
        }

   \institute{Key Laboratory of Dark Matter and Space Astronomy, Purple Mountain Observatory, Chinese Academy of Sciences, Nanjing, People's Republic of China
   \and School of Astronomy and Space Science, University of Science and Technology of China, Hefei, People's Republic of China}

   \date{Accepted 23 Sep, 2026}

   \titlerunning{BHs in red-sequence galaxies: Not source of dark energy, but remnants of LRDs}
   \authorrunning{Lei et al.}

  \abstract
   {The nature of dark energy remains one of the most profound mysteries in modern cosmology. One intriguing proposal is that black holes (BHs) could be the astrophysical source of dark energy through a cosmological coupling mechanism, and strong evidence has been claimed via analyzing the growth of the black hole masses in the red-sequence elliptical galaxies at redshifts $\leq 2.5$. In this Letter, we show with a group of very high-redshift active galactic nuclei detected by the James Webb Space Telescope (JWST) in the red passively evolving galaxies that the possibility that BHs are the astrophysical source of dark energy has been rejected at a confidence level of 8.1 $\sigma$. Moreover, our analysis suggests that the little red dots recently discovered by JWST are consistent with being the progenitors of red-sequence elliptical galaxies hosting the relatively low-mass BHs at redshifts of $\sim 0.7-2.5$. This offers an alternative explanation without the need for black hole cosmological coupling.}


   \maketitle

\section{Introduction}\label{sec:intro}

Dark energy is thought to be the driving force behind the accelerated expansion of the Universe \citep{1998AJ....116.1009R,1999ApJ...517..565P}, but its  nature remains elusive \citep{Frieman:2008sn,Li:2011sd,2009ARNPS..59..397C}. Various models have been proposed to account for the dark energy, including the cosmological constant \citep{1998AJ....116.1009R,1999ApJ...517..565P}, the scalar field \citep{Caldwell:1997ii,Ferreira:1997hj,2025PhRvD.112f3508S}, quintessence \citep{2005PhRvL..95n1301C,Tsujikawa:2013fta}, phantom \citep{2003PhRvL..91g1301C}, quintom \citep{Cai:2009zp,2010MPLA...25..909Q}, modified gravity \citep{Cai:2015emx}, coupled dark energy scenarios \citep{Amendola:2002bs,Amendola:2003eq}, quantum gravity\citep{Li:2011sd}, holographic dark energy \citep{2004PhLB..603....1L,2004PhLB..594...13H,2007NuPhB.774..256C,Li:2011sd}, cosmological back-reaction \citep{2007PhRvD..76l3004M,2008PhRvD..77b3003M,2007JCAP...12..017B,2008JCAP...06..021B}, Chaplygin gas \citep{2001PhLB..511..265K,2002PhRvD..65d3507A,2025ApJ...993..155W}, viscous fluid \citep{2005PhRvD..72b3003N,2006PhLB..633....1R,Meng:2005jy},
and particles with exotic properties \citep{DeDeo:2004vi,2008PhRvD..77d5015T,2009PhRvD..80h3513S,2011IJMPD..20.1235D,2025EPJC...85..449L,2011PhRvD..84f3511J,2010PhRvD..82b3504B}.

A recent proposal, invoking a nonstandard coupling between local compact objects and global cosmic expansion, suggests that BHs, particularly stellar remnants, might be the astrophysical origin of dark energy \citep{2019ApJ...882...19C,2020ApJ...889..115C,2020ApJ...900...57C,2023ApJ...943..133F,Farrah:2023opk}. This mechanism can be realized through non-singular BHs embedded in the cosmological background \citep{Faraoni:2024ghi,Cadoni:2023lqe,Faraoni:2023hin,Cadoni:2024rri,Almatwi:2024utb,2026EPJC...86..288W} (see, however,
\cite{Avelino:2023rac,Mistele:2023fds,Gaur:2023hmk,Dahal:2023hzo}). According to this hypothesis, the mass of BHs would grow with the cosmic scale factor \(a\) as \(M_{\rm BH}(a) = M_{\rm BH}(a_i) \times \left( a/{a_i} \right)^k\), where \(k\) is the cosmological coupling strength \citep{Farrah:2023opk}. If  $k\sim 3$, the BHs contribute as a cosmological dark energy species. The strength $k=0$ instead represents the absence of coupling between BHs and the accelerated expansion of the universe. A preferential growth channel for supermassive BHs in red-sequence elliptical galaxies, identified by \cite{2023ApJ...943..133F}, {provided preliminary evidence} of a strong cosmological coupling with strength \(k = 3.09 \pm 0.76\) \citep{Farrah:2023opk}.
Moreover, the astrophysical BHs produced during the process of star formation can meet the cosmological dark energy density proportion of $\Omega_{\Lambda} = 0.68$ through cosmological coupling, without violating the existing limits of the abundance of the massive compact halo objects (MACHOs) \citep{Farrah:2023opk}.

As the astrophysical origin of dark energy, cosmologically coupled BHs exhibit a unique mass-growth mechanism that is independent of accretion and mergers.
For instance, the cosmologically coupled mass growth of BHs can solve the low-mass gap problem \citep{Gao:2023keg,2021ApJ...921L..22C}
and provide a time-evolving dark energy source that naturally explains the recent measurements by the Dark Energy Spectroscopic Instrument (DESI) \citep{Croker:2024jfg}.
The cosmologically coupled BHs can be used to explain higher-mass objects as observed in the LIGO-Virgo mass distribution \citep{2021ApJ...921L..22C}. However, this explanation is not particularly necessary when the existence of a sub-population of hierarchical mergers has been confirmed \citep{2022ApJ...941L..39W,2024PhRvL.133e1401L,2025ApJ...987...65L,2026arXiv260701121L}. 
The cosmologically coupled BHs can fit DESI data well with a peaked positively summed neutrino mass $\Sigma m_\nu = 0.106$ \citep{2025PhRvL.135h1003A}. These objects can provide anisotropic dark energy \citep{Chan:2011ayt}, which has also received widespread attention \citep{Zhang:2023neo,McConville:2023xav}.

Dedicated tests have been carried out, but the conclusions are controversial. 
\cite{Cadoni:2023lum} reached the same conclusion  as  \cite{2023ApJ...943..133F} based on similar data.  The gravitational-wave background (GWB) from SMBHs measured by pulsar timing arrays favors the strong coupling strength $k\approx3$ \citep{Lacy:2023kbb,Calza:2024qxn}.
On the other hand, \cite{Andrae:2023wge} found that the probability of a non-cosmological coupling was 77\% based on Gaia observations.
The high-redshift active galactic nuclei (AGNs) observed by JWST tend to suggest an absence of cosmological coupling \citep{Lei:2023mke}.
The minimum formation mass of the BHs in the GWTC-3 observations disfavors the cosmological coupling model at a level of $2\sigma$ \citep{Amendola:2023ays}.  
A non-coupling BH population is favored with stellar mass BHs in X-ray binaries \citep{Mlinar:2023fkk}. 
The mass function of BHs in the globular cluster NGC 3201 excluded a cosmological coupling of BHs at $3.8\sigma$ \citep{Rodriguez:2023gaa}. {Notably, the exclusions of \cite{Rodriguez:2023gaa,Andrae:2023wge} were based on the assumption of the Tolman–Oppenheimer–Volkoff (TOV) limit as a lower bound on the BH mass, which may not apply to vacuum-energy interiors.}
Therefore, a decisive constraint at a confidence level above $5\sigma$ is still lacking. 

In this Letter, we take the sample of high-redshift AGNs detected by the JWST to assess whether BHs are viable as the source of dark energy through cosmological coupling. The cosmological parameters we used include $H_{0}=67.36\, \rm km\,  s^{-1} \, Mpc^{-1}$, $\Omega_{m}=0.3135$, and $\Omega_{\Lambda}=0.6847$ \citep{2020A&A...641A...6P}. Our main finding is that this possibility is rejected at a confidence level of {$\sim 8\,\sigma$}. We also show that the little red dots (LRDs) recently discovered by JWST can naturally evolve into red-sequence elliptical galaxies at redshifts of $\sim 0.7-2.5$, which further weakens the black hole dark energy hypothesis.

\section{Data and analysis}
\label{sec:data}

The mass of BHs coupled with cosmology evolves significantly with redshift. In this theory, the mass of BHs in AGNs grows noticeably, regardless of the host galaxy over time, while the stellar mass of early-type galaxies does not change much during the evolution \citep{2020ApJ...889..115C,Farrah:2023opk}. Therefore, to examine the mass-growth patterns caused by the cosmological coupling of BHs in AGNs, we selected samples of sufficiently old host galaxies \citep{2023ApJ...943..133F}. 

We selected luminous AGNs residing in early-type host galaxies with high stellar mass that were observed by JWST to avoid biases from the low-luminosity population. 
Our AGN sample of early-type host galaxies is summarized in Table \ref{Tab:data}.

\begin{figure}[ht!]
\centering
 \includegraphics[width=0.99\linewidth]{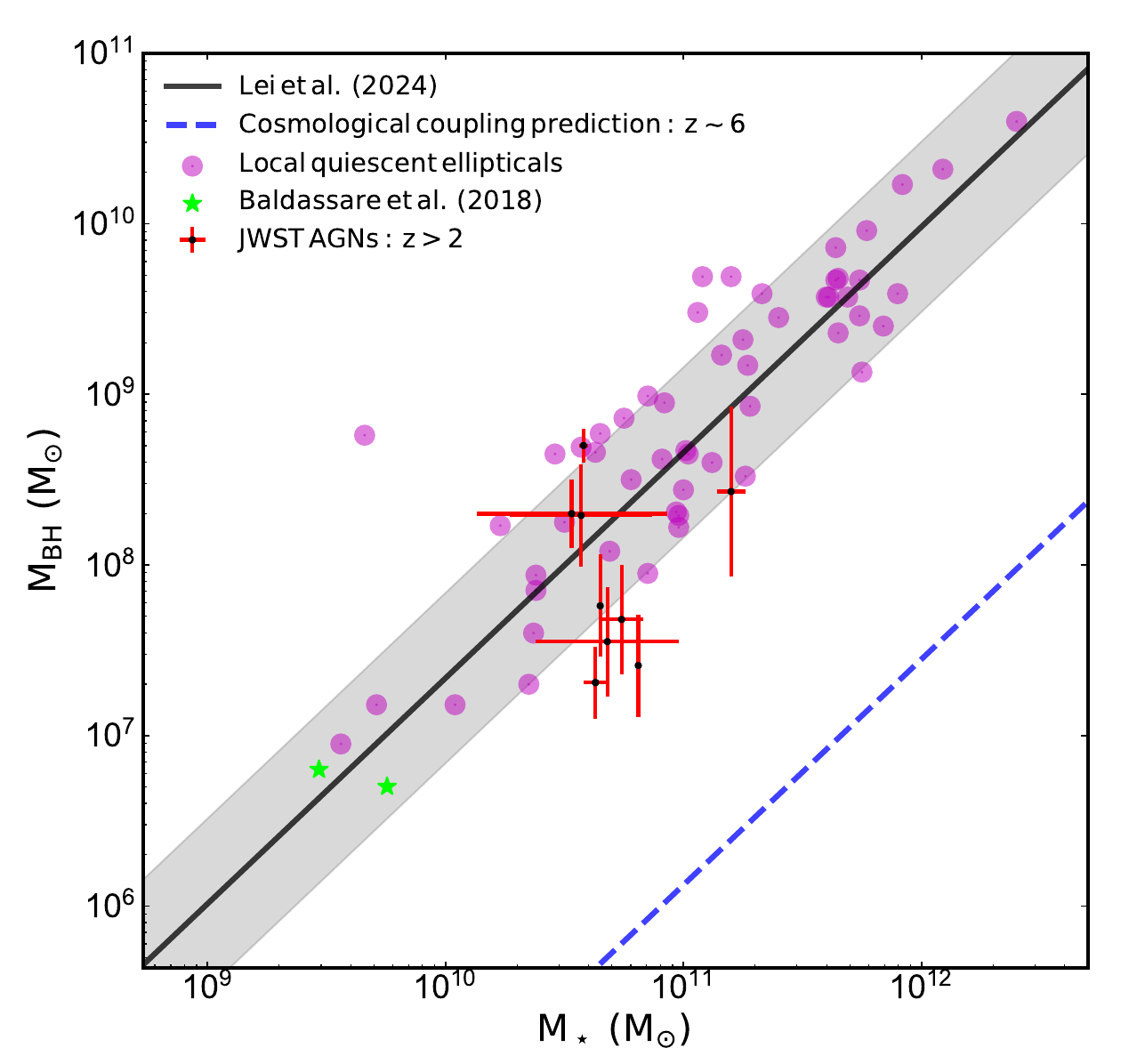}
 \caption{ Our selected AGNs and the local sample. The black line shows the $M_{\star}-M_{\rm BH}$ fundamental plane relation of local elliptical host AGNs. The data points with error bars show the nine selected AGNs. 
The purple circles show local quiescent elliptical AGNs from \cite{2023ApJ...943..133F}. The green stars show two local AGNs in dwarf early-type galaxies taken from \cite{Ben-Ami:2017gwo}. The dashed blue line represents the predicted relation when the cosmological coupling strength is set to $k=3$. }
 \label{fig:data}
\end{figure}

Our selected sample is shown in Fig. \ref{fig:data}.
The $M_{\star}$ - $M_{\rm BH}$ fundamental plane is a well-established empirical relation \citep{Kormendy:2013dxa,2015ApJ...813...82R,2023NatAs...7.1376Z}. 
For the AGNs in early-type host galaxies, we have 
\cite{Lei:2023mke}
\begin{equation}
    \log_{10}( \frac{M_{\rm BH}}{M_{\odot}} )
    = 8.66_{-0.01-0.5}^{+0.01+0.5}+1.32_{-0.01}^{+0.01}\times \log_{10} ( \frac{M_{\star}}{10^{11}\ M_{\odot}} ),
    \label{eq:fund}
 \end{equation}
where \(M_{\text{BH}}\) is the black hole mass, and \(M_{\star}\) is the stellar mass of the host galaxy. The errors are the $1\sigma$ standard deviations found in the fitting, and the $0.5\ \rm dex$ refers to the intrinsic scatter of the local sample shown in the gray region in Fig. \ref{fig:data}.
If there is a cosmological coupling between BHs and dark energy, we would expect the intercept of this relation to evolve with redshift as \(M_{\rm BH}(z_i) = M_{\rm BH}(z_0) \times \left( 1+{z_i} \right)^{-k}\).

To estimate the parameter $k$, we used the relation of
{\begin{equation}
\begin{array}{c}
\mu_{i}(k)=1.32\times(\log_{10}(\frac{M_{\star,i}}{M_{\odot}}) - B_{\star})-k\times \log_{10}(1+z_{i})-5.86,
    \label{eq:k}
\end{array}
\end{equation}
whose likelihood reads \citep{Lei:2023mke}
\begin{equation}
\mathcal{L}\left(d \vert k\right) = \prod_i \frac{\exp \left\{ -\frac{\left[\mu_{i}\left(k\right)-\left( \log_{10}\left(\frac{M_{{\rm BH},i}}{M_{\odot}}\right)-B_{\rm BH}\right) \right]^2}{2(\sigma_{i})^2} \right\}}{\sigma_{i}\,\sqrt{2\pi}} ,
\label{eq:7}
\end{equation}
where $M_{{\rm BH},i}$ and $\sigma_{i}$ are the observed BH mass and the error including the uncertainties of $M_{{\rm BH,}i}$ and $M_{\star,i}$, respectively. 
The total error  includes the measured BH mass error ($\sigma_{{\rm BH,}i}$), the conversion error ($\sigma_{{\rm conv},i}$) from the stellar mass and the fundamental plane intrinsic scatter $\sigma_{\rm intrinsic}=0.51\;\rm dex$: 
$\sigma^2_{{\rm tot},i}=\sigma^2_{{\rm BH},i}+\sigma^2_{{\rm conv},i}+\sigma^2_{\rm intrinsic}$.
The errors in our data are in the logarithmic space. Here, $z_i$ denotes the cosmological redshift of the source. To account for potential biases, we introduced the parameters \( B_{\star} \) and \( B_{\rm BH} \), corresponding to the measurement biases in stellar mass and black hole mass, respectively (see also Appendix \ref{app:b}). }

\section{Result and discussion}
\label{result}

\begin{figure}[ht!]
\centering
 \includegraphics[width=0.99\linewidth]{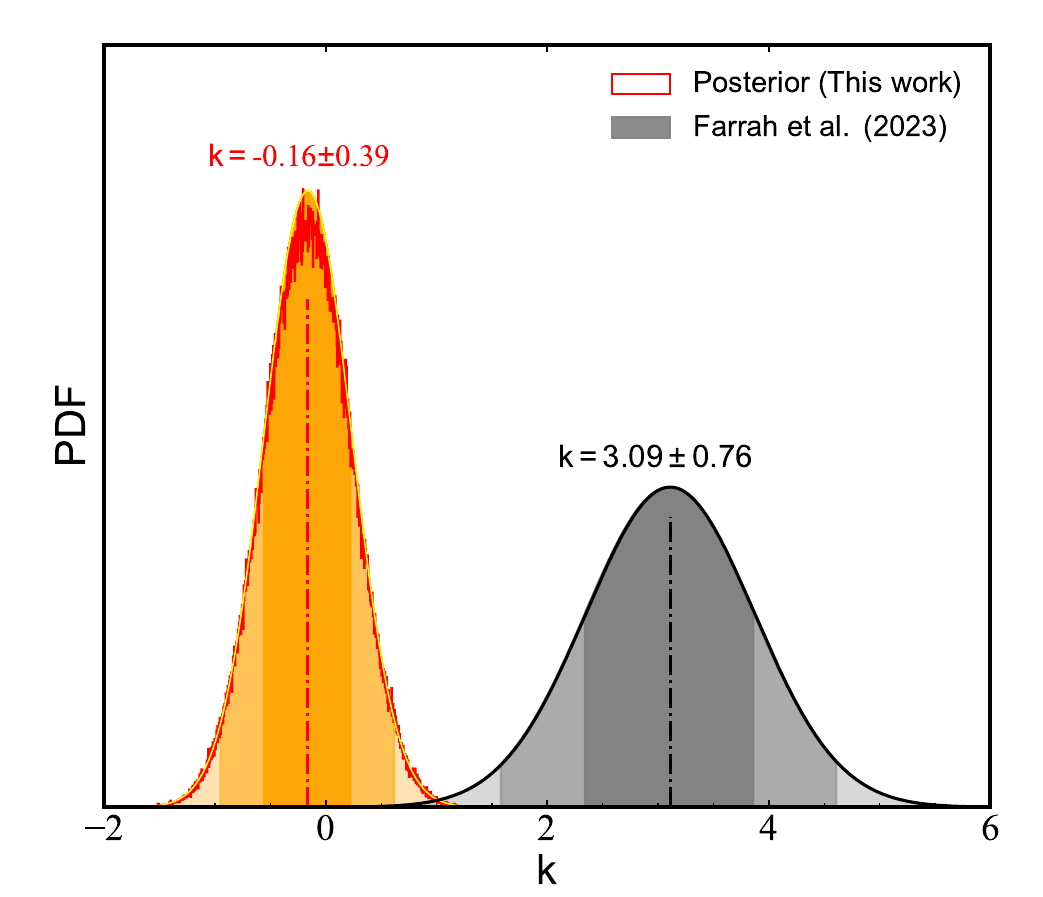}
 \caption{Posterior of the cosmological coupling strength $k$. }
 \label{fig:post}
\end{figure}

As shown in Fig. \ref{fig:post}, the constraints of the coupling strength parameter \(k\) are found to be 
\begin{equation}
k = -0.16 \pm 0.39\,\rm  (68\%\, confidence\, interval).
\end{equation}
Therefore, the deviation from the value of \(k = 3\) required for BHs to serve as dark energy indicates a confidence level of $8.1\, \sigma$. This is remarkably consistent with the prediction of the non-cosmologically coupled BHs ($k=0$).

We conclude that 
the JWST high-redshift AGNs have convincingly ruled out the possibility that BHs are the source of dark energy. 
As shown below, these red-sequence objects are likely the descendants of LRDs, a specific group of recently discovered JWST sources in the high-redshift universe.

The LRDs are generally characterized by their V-shaped spectrum, broad emission lines, and compact morphologies compared to bright AGNs \citep{Matthee:2023utn,2025ApJ...978...92L,2023ApJ...956...61A,2025A&A...701A.168D,2026ApJ...999...70L,2025ApJ...989L..30S}. LRDs might host intermediate-mass BHs accreting at high or super-Eddington rates (e.g., \cite{2026arXiv260630711N,2026arXiv260317667M,2026ApJ..1005L..37T,2025arXiv250316596N,2026Natur.649..574R,2026ApJ...999..183U,2026ApJ...996..129G}), which naturally explains their observed X-ray weakness \citep{2026arXiv260222386M,2025PASJ...77..811I,2025ApJ...988L..22I,2024ApJ...976...96P}. 

\begin{figure}[ht!]
\centering
 \includegraphics[width=0.99\linewidth]{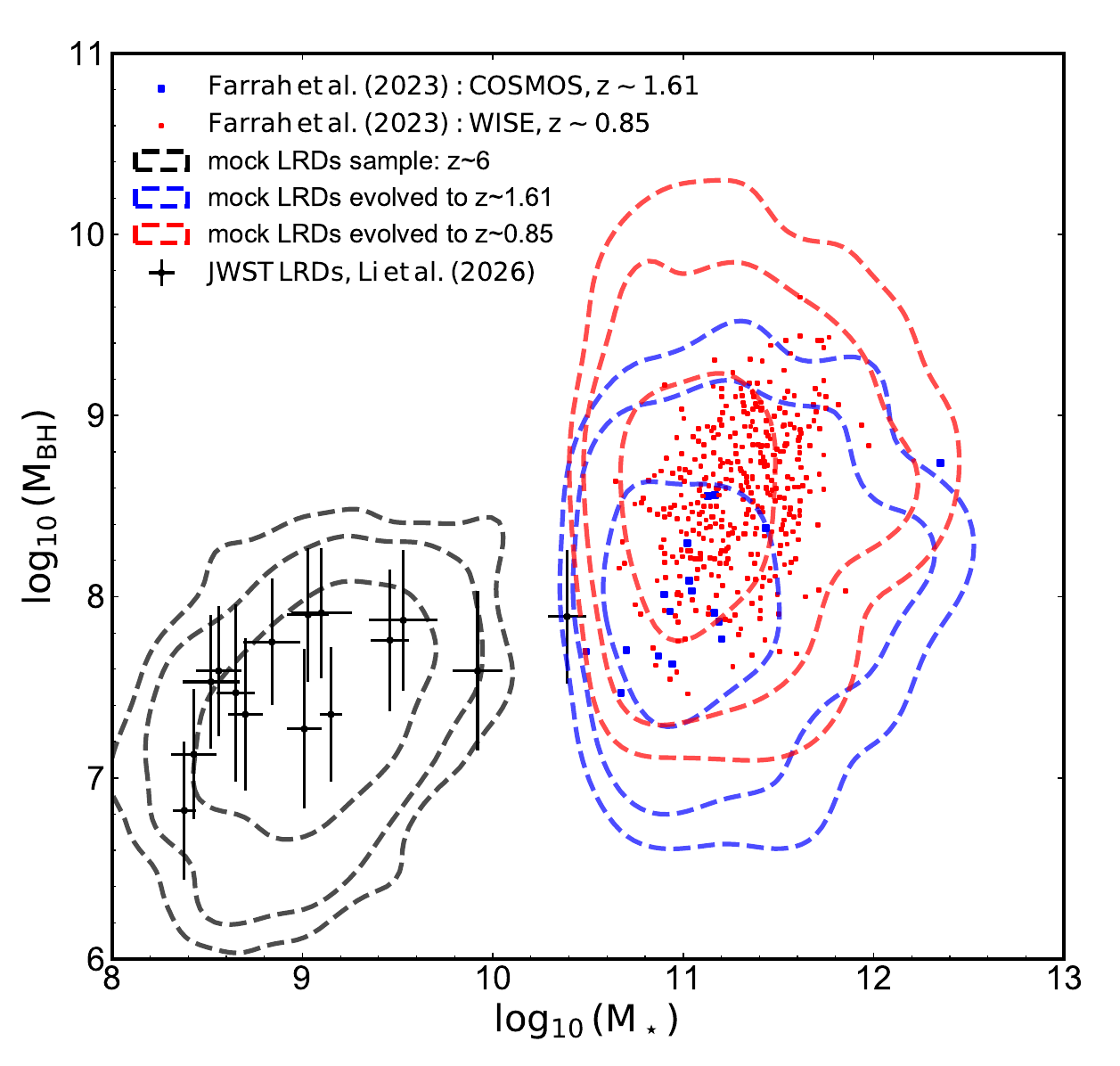}
 \caption{ Evolution history of the JWST LRDs. It is consistent with the evolution of the sample of AGNs with red-sequence elliptical hosts. The selected $z\sim 0.85$ and $z\sim 1.61$ AGNs from \cite{Farrah:2023opk} are shown as red and blue points. The dashed contours in black, blue, and red show our simulated LRDs at redshift $z\sim6$, $z\sim 1.61$, and $z\sim 0.85$. The three contours correspond to sample density levels of 0.05, 0.1, and 0.3, respectively. The black points with error bars are JWST LRDs adopted from \cite{2026ApJ...999...70L}.  }
 \label{fig:LRD}
\end{figure}

The LRDs are thought to have a strong star formation potential \citep{Matthee:2023utn,2025NatCo..16.9830T,2026A&A...709L..11D}, and the peak in the number density redshift distribution of this group of targets is in the redshift range of $5-8$ \citep{2026ApJ..1000...59M,2025ApJ...989L..19P}, which is significantly different from the peak in the density distribution of ordinary bright AGNs (at redshift of $1-3$, \cite{2026ApJ..1000...59M}). As shown below, this provides a viable explanation for the evolution of the specific black hole population reported in \cite{Farrah:2023opk}. 

To explore whether LRDs might provide a natural astrophysical explanation for the sample used in \cite{Farrah:2023opk}, we conducted simulations of the LRD evolution in the $M_{\rm BH}-M_{\star}$ plane, as illustrated in Fig.~\ref{fig:LRD}. Using the sample of LRDs observed by JWST at $z \approx 6$ \citep{2026ApJ...999...70L}, we generated 1000 {mock} sources whose stellar masses and black hole masses followed a two-dimensional log-normal distribution consistent with the observed properties of this population. 

We modeled the LRD mass growth from $z\sim 6$ to $z\sim 1.61$ assuming constant SFRs and a baseline constant BH accretion rate. This baseline accretion rate is directly based on recent empirical constraints of LRDs \citep{2026ApJ...999...70L} (detailed in Appendix \ref{app:c}). From $z\sim 1.61$ to $z\sim 0.85$, we assumed that the host galaxies quench ($\rm{SFR}=0\, M_{\odot}\, yr^{-1}$ with a 0.8 dex scatter) to match the red-sequence ellipticals observed by \cite{Farrah:2023opk}. Crucially, despite this quenching of the star formation, frequent galaxy mergers as inferred from strong LRD clustering \citep{2026ApJ...999...31Z,2024arXiv241214246T} can sustain continuous central gas inflows, which physically justifies our assumption of a maintained accretion rate even after quenching.

Furthermore, we recognize that the true accretion histories of LRDs might be more complex. Recent interpretations suggested that LRDs could host lower-mass BHs accreting at super-Eddington rates \citep{2026arXiv260630711N,2026arXiv260317667M,2026ApJ..1005L..37T,2025arXiv250316596N,2026Natur.649..574R,2026ApJ...999..183U,2026ApJ...996..129G,2025ApJ...980L..34L,2026ApJ..1005..150L}. To test the robustness of our baseline model, we simulated a suite of alternative accretion scenarios (detailed in Appendix \ref{app:c}), including episodic super-Eddington accretion, merger-driven bursts, and exponentially declining models. Remarkably, all these alternative physical scenarios yielded evolutionary outcomes that were consistent with our constant accretion baseline. Our simulation demonstrates that the LRDs currently observed by JWST can naturally explain the BHs in the red sequence at redshifts of $0.7-2.5$ \citep{Farrah:2023opk}, without the need to introduce black hole cosmological coupling.

\begin{acknowledgements}
We thank Duncan Farrah, Kevin S. Croker, Lei Zu, Guan-Wen Yuan, Zhen-Bo Su, Wen-Ke Ren, Chi Zhang, and Da-Ming Wei for helpful discussions. 
The authors were supported in part by the National Key Research and Development Program of China (No. 2022YFF0503301),  NSFC under
grants of No.12233011 and No.12373002, and the New Corner stone Science Foundation
 through the XPLORER PRIZE.
\end{acknowledgements}

%
\bibliographystyle{aa} 
\bibliography{refs} 
\begin{appendix}




\twocolumn

\section{Selected AGN sample in JWST observation}

JWST is well suitable to measure high redshift objects, and these data have been widely adopted to test the models of dark energy or dark matter \citep{2023A&A...672A..71M,Wang:2023xmm,2025A&A...699A..59S,2025PhRvL.134g1002J,2025PhRvL.134g1003R,Lei:2025ooq,Wang:2024hce,2022PhRvD.106i5025B,2025PhRvL.134e1004Y,2026A&A...707A.300M}. The high-resolution spectra of JWST have been widely adopted to search for AGNs, as their significant broad emission line features can be used to calculate black hole mass \citep{2023Natur.621...51D,2026ApJ...997..250Z,2026MNRAS.546ag086J}. The JWST's highly sensitive photometric camera can also provide high quality images to separate the host galaxies of AGNs \citep{2026ApJ...999...70L,2024ApJ...966..176Y}.

Among the numerous AGNs observed by JWST, more than 10 host galaxy stellar masses exceed $4\times 10^{10}$ solar masses \citep{2023Natur.621...51D,2024ApJ...975..178K,2026ApJ...997..250Z,2026MNRAS.546ag086J,2026ApJ...999...70L}, making these samples suitable for testing cosmological coupled black hole theories.

To avoid the effects of accretion growth of BHs, star formation, and other factors on the redshift evolution of the relationship between supermassive black hole mass and host galaxy stellar mass, the selected sample meets the following conditions \citep{2023ApJ...943..133F,Farrah:2023opk,Lei:2023mke}:
\begin{itemize}
\item Redshift: We select the AGNs at high redshift $z>2.0$. As listed in Table~\ref{Tab:data}, we selected the sample in a range of $2.0<z<7$. 

\item Dust extinction: 
We select the blue AGNs with {${E(B-V)} < 0.21$}, excluding the sample with relatively high BH accretion rates and star formation rates.

\item Host type: Following \cite{2023ApJ...943..133F,Farrah:2023opk},  we select the early-type host galaxies without structures like shells, spiral arms or disks in their images. It can remove the sample with complex evolutionary history and star formation history.

\item Stellar mass ($M_{\star}$): Similar to \cite{2023ApJ...943..133F,Farrah:2023opk}, we select the sample with host
stellar masses higher than {$3.3\times 10^{10}\rm M_{\odot}$}, ensuring their passive evolution and small selection biases.

\item Star formation rate (SFR): Adopting a stricter criterion than \cite{Farrah:2023opk} (i.e., $\mathrm{SFR} < \psi_{\mathrm{SFR}}/5$), we select the AGN host galaxies that satisfy $\mathrm{SFR} < \psi_{\mathrm{SFR}}/5 - \delta_{\mathrm{SFR}}$, where $\psi_{\mathrm{SFR}}$ is the main sequence model \citep{Speagle:2014loa,Heintz:2022qql} and $\delta_{\rm SFR}$ is the intrinsic scatter. It ensures the sample evolves passively.

\end{itemize}

The ${\rm SFR} - M_{\star}$ main sequence relationship at $z\sim 0-6$ is adopted from \cite{Speagle:2014loa}, while for {the redshift range of} $z\sim 6-10$ we use the relationship suggested by the JWST  data \citep{Heintz:2022qql}.
The SFRs of AGN host galaxies are estimated using H$\alpha$ narrow emission line.

Our total AGN sample of early-type host galaxies is summarized in Table \ref{Tab:data}. 

\begin{table*}[hbtp]
\centering
\caption{The selected JWST AGN sample in this work.}
\label{Tab:data}
\begin{tabular}{ccllccc}
\hline\hline
ID & $z$ & E(B-V)  & SFR ($\rm \frac{M_{\odot}}{yr}$) & $\log_{10}{(\frac{M_{\star}}{M_{\odot}})}$ &  $\log_{10}{(\frac{M_{\rm BH}}{M_{\odot}})}$ & References  \\ \hline     
{NIRSpec-8777}   & {4.66}  & {0.001}   & {$0.03\pm 0.01$\textsuperscript{b}}       & {$10.81\pm0.01$} &  {$7.41\pm0.30$}  & {\cite{2025MNRAS.537.3453B,2025MNRAS.539..557B}, this work \textsuperscript{d}} \\
{ZF-UDS-3651}   & {3.81}  & {0.10}   & {$0.27\pm 0.07$\textsuperscript{b}}       & {$10.65\pm0.01$} &  {$7.76\pm0.30$}  & {\cite{2024NatSR..14.3724N}, this work \textsuperscript{d}} \\
GN-72127   & 4.13  & 0.11   & $5.7\pm 1.2$\textsuperscript{b}       & $10.63\pm0.05$ &  $7.31\pm0.21$  & \cite{2024ApJ...975..178K}  \\
GN-9994014 & 5.32  & {$0.06$ \textsuperscript{a}} & $13.2\pm 3.2$\textsuperscript{b}      & $10.68\pm0.30$ &  $7.55\pm0.30$  & \cite{2026ApJ...997..250Z} \\
{GN-1087388} & {5.54}  & {$0.06$ \textsuperscript{a}} & {$62.4\pm 1.4$ \textsuperscript{b}}     & {$10.57\pm0.30$} &  {$8.29\pm0.30$}  & \cite{2026ApJ...997..250Z} \\
GS-204851  & 5.48  & 0.02   & $21.4\pm 8.1$\textsuperscript{b}      & $10.74\pm0.09$ &  $7.68\pm0.32$  &  \cite{Matthee:2023utn,2026MNRAS.546ag086J} \\
{J2255+0251} & {6.33}   & {0.17}   & {$38.1\pm7.5$ \textsuperscript{b,c}}             & {$10.53\pm0.44$} &  {$8.30\pm0.20$}  & {\cite{2023Natur.621...51D}}  \\
COS-XQG1   & 2.09  & 0.21   & $1.5\pm 0.2$\textsuperscript{b}       & $11.20\pm0.06$ &  $8.43\pm0.50$  & \cite{2025MNRAS.538.1501I}  \\
{GS-9209}   & {4.66}  & {0.01}   & {$1.9\pm 0.1$\textsuperscript{b}}       & {$10.58\pm0.02$} &  {$8.70\pm0.10$}  & \cite{2023Natur.619..716C}  \\
\hline\hline
\end{tabular}
\begin{tablenotes}
 \item Notes:\\
 \textsuperscript{a} {The E(B–V) was estimated using the extinction law from \cite{2000ApJ...533..682C} and the UV slope provided by  \cite{2026ApJ...997..250Z}.}
 \item[b] \textsuperscript{b} The SFR is estimated with the luminosity of the narrow H$\alpha$ line. The luminosity of the H$\alpha_{narrow}$ line of GS-204851 is from the result fitted in  \cite{Matthee:2023utn}.
 \item[c] \textsuperscript{c} The SFR values are estimated with the  luminosity of the narrow H$\alpha$ line. The narrow H$\alpha$ line luminosity is converted from the narrow H$\beta$ line luminosity by a factor $L_{H\alpha, \rm broad}/L_{H\alpha, \rm narrow}=3.05$ (a statistical result of $z>5$ AGNs \citep{2026MNRAS.546ag086J}).
 \item[d] \textsuperscript{d} {We calculated E(B-V) from fitting the continuum spectrum of galaxy. We also fitted the broad and narrow H$\alpha$ emission lines for calculating the SFR and $M_{\rm BH}$.  For details, see the Appendix \ref{app:a}.}
\end{tablenotes}
\end{table*}

\subsection{Spectral analysis of two AGNs in quiescent host galaxies}\label{app:a}

{In \cite{2025MNRAS.537.3453B,2025MNRAS.539..557B} and \cite{2024NatSR..14.3724N}, the spectra of two quiescent galaxies, NIRSpec-8777 and ZF-UDS-3651, have been discussed, respectively. Both sources were found to exhibit strong H$\alpha$ emission lines, which were speculated to be likely caused by AGN-driven intense emission. The spectra of these two sources exhibit strong Balmer breaks and absorption lines, indicating the presence of an old stellar population and suggesting that they are passively evolving galaxies, consistent with our initial selection criteria. However, we need to quantitatively determine their E(B–V), SMBH masses and SFR values.
The data for these two sources are publicly available in the Barbara A. Mikulski Archive for Space Telescopes\footnote{MAST: \url{https://mast.stsci.edu/portal/Mashup/Clients/Mast/Portal.html}}. We downloaded the low- and medium-resolution JWST/NIRSpec spectral data for both sources accordingly.}

{To estimate the E(B-V) values for these two sources, we performed stellar component fitting to the low-resolution spectra, as the continuum coverage from approximately 2500 to 8500 $\AA$ in the rest frame provides strong constraints on the E(B-V) parameter. In the fitting process of stellar continuum, we masked the spectral region around the H$\alpha$ emission line to fit the continuum. For ZF-UDS-3651, we also masked the region around the prominent [O III] emission line during the stellar continuum fitting. }

{During the spectral fitting process, we modeled the stellar continuum using a continuity non-parametric star formation history reconstruction \cite{2019ApJ...876....3L}, which was implemented with the Bagpipes code \cite{2018MNRAS.480.4379C}. For the continuity non-parametric priors, we adopted 10 bins with edges corresponding to lookback times of 0, 1, 2, 5, 10, 50, 100, 250, 500, and 1000 Myr. A log-uniform prior was applied to the total stellar mass of the galaxy. Additionally, a log-uniform prior was adopted for metallicity. For the extinction model, we used the \citet{2000ApJ...533..682C} dust extinction law with a uniform prior on $A_V$ in the range [0, 4].
}

\begin{figure}[htbp]
\centering
 \includegraphics[width=0.99\linewidth]{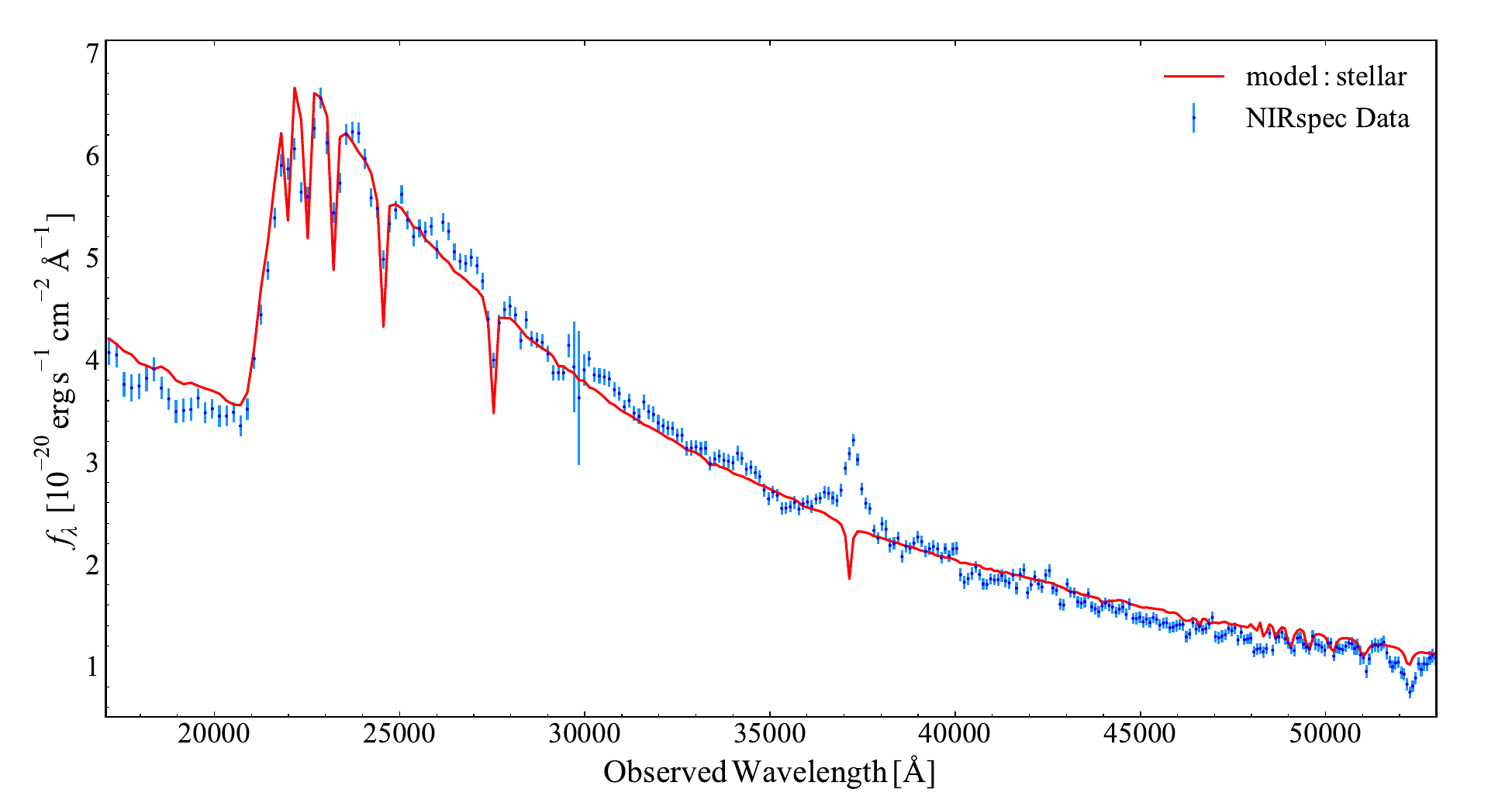}
 \includegraphics[width=0.99\linewidth]{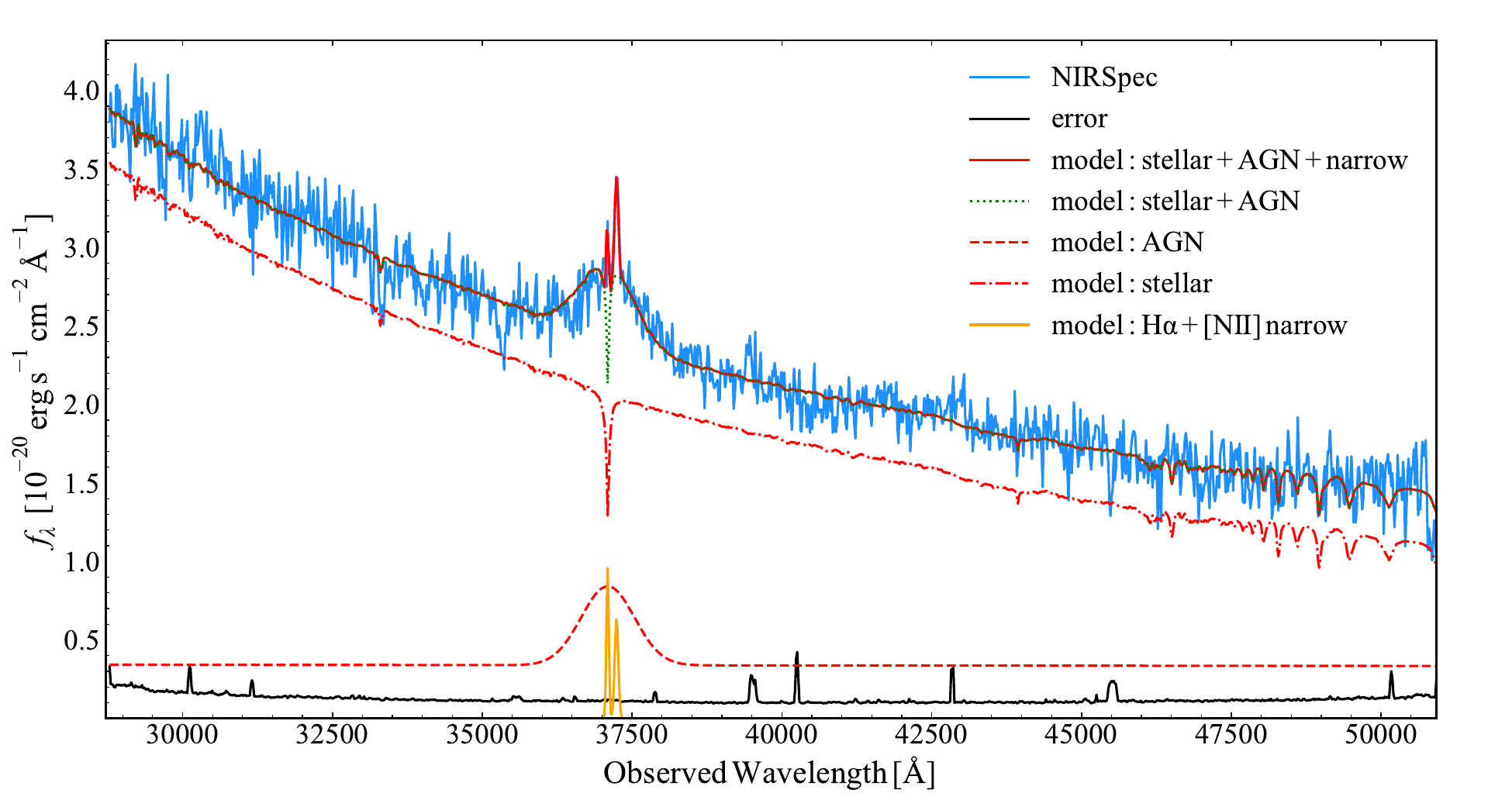}
 \caption{{JWST spectrum of the quiescent galaxy NIRSpec-8777 hosting an AGN.  
Top panel: Low-resolution spectrum of galaxy NIRSpec-8777 (blue data points with error bars) and the best-fit stellar spectral model (solid red line).  
Bottom panel: Medium-resolution spectrum around the H$\alpha$ emission line (blue solid line: observed values; black solid line: observational uncertainties) and the fitting results (solid red line: total model flux; red dot-dashed line: stellar component model; red dashed line: AGN component model flux including broad H$\alpha$ emission and continuum; solid orange line: narrow emission line model flux for H$\alpha$ and [N II]).} }
 \label{fig:8777}
\end{figure}

\begin{figure}[htbp]
\centering
 \includegraphics[width=0.99\linewidth]{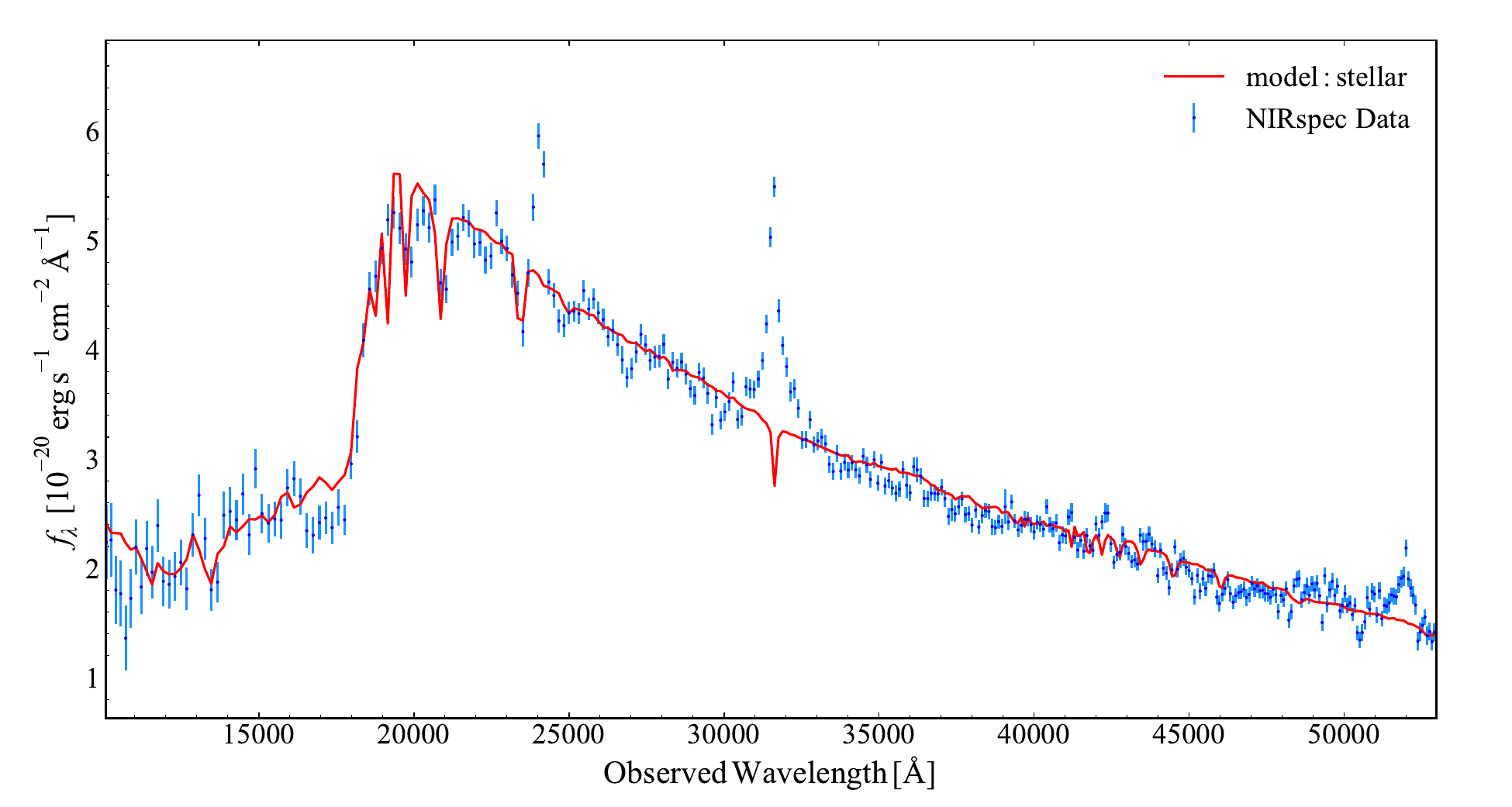}
 \includegraphics[width=0.99\linewidth]{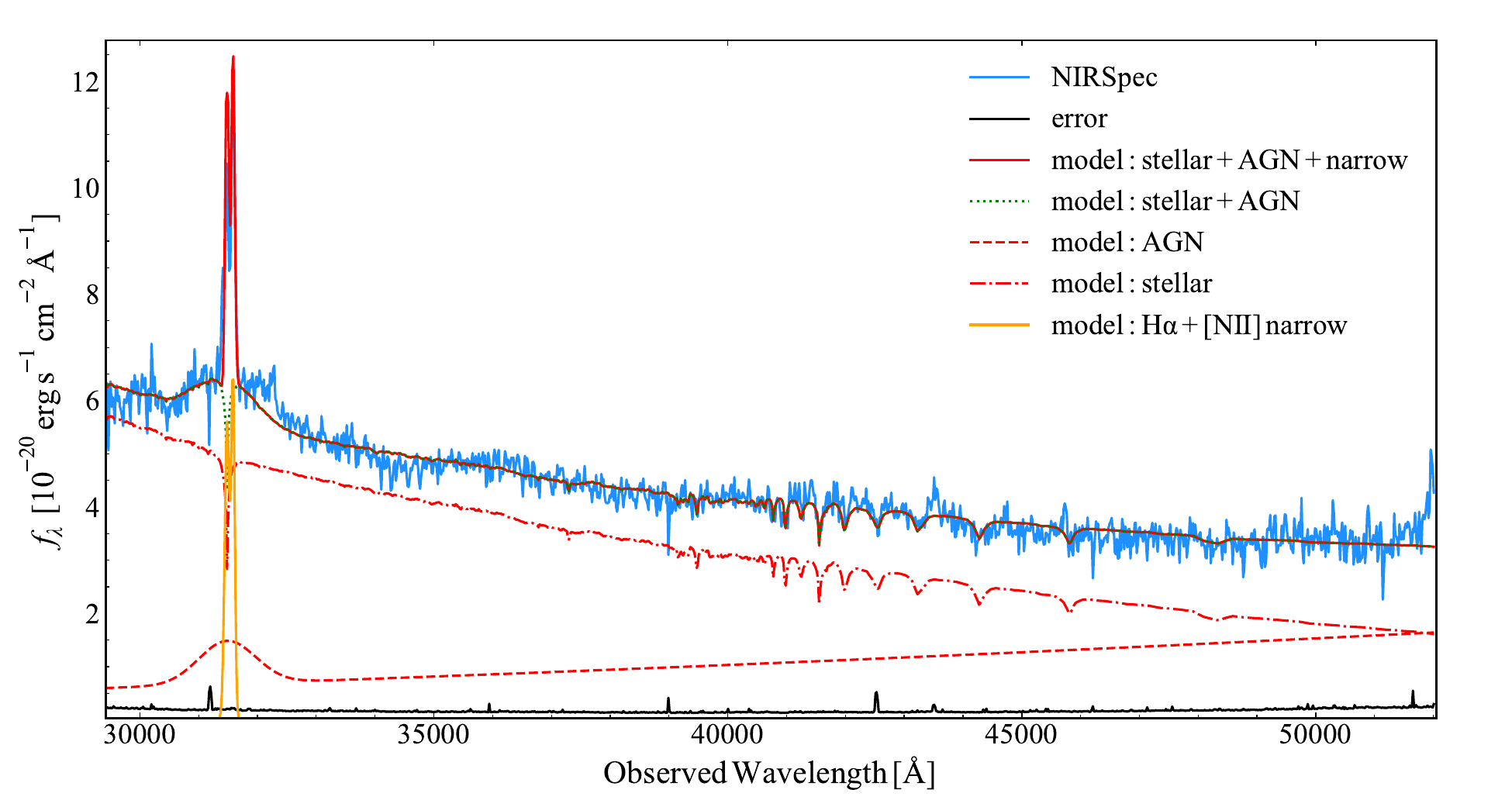}
 \caption{{Same as Fig. \ref{fig:8777}, but for ZF-UDS-3651.} }
 \label{fig:3651}
\end{figure}

{As shown in the upper panels of Figs. \ref{fig:8777} and \ref{fig:3651}, our model provides an best fit to the stellar continuum of both galaxies NIRSpec-8777 and ZF-UDS-3651. The best-fit extinction parameters for the continua of these two sources are \( A_V = 0.004 \pm 0.004 \) and \( A_V = 0.40 \pm 0.02 \), respectively. After converting via the relation \( E(B-V) = A_V / 4.05 \), the corresponding reddening values are \( E(B-V) = 0.001 \) and \( E(B-V) = 0.10 \), indicating that both sources are dust-poor passively evolving galaxies. 
}

{To more accurately estimate the SFR and black hole mass of these two sources, we used Bagpipes \cite{2018MNRAS.480.4379C} to fit the JWST medium-resolution spectra (with a mean spectral resolution of \( R = 1000 \)) around the broad H$\alpha$ emission line. For the medium-resolution spectra, we employed the AGN module in Bagpipes \cite{2018MNRAS.480.4379C} to model both the AGN continuum contribution and the broad H$\alpha$ emission line. The stellar continuum was fitted using the same parameter ranges and priors as those adopted in the continuum fitting described above. The broad H$\alpha$ emission line was modeled with a Gaussian profile, with the flux normalization factor \( n_{H\alpha} \) and the line width \( \sigma_{H\alpha} \) treated as free parameters. The key parameter for black hole mass estimation, \( \sigma_{H\alpha} \), was assigned a uniform prior over the range 1000–6000 km/s. We used a log-uniform prior over \(n_{H\alpha} \in [10^{-22}, 10^{-17}] \, \rm erg/s/cm^2/\AA\) for the prior of the flux normalization. 
}

{We note that when using the Bagpipes code \cite{2018MNRAS.480.4379C} to model the spectrum through a combination of stellar and AGN components, significant narrow emission lines of H$\alpha$ and [N II] remained present in the data residuals. Thus, when fitting the broad H$\alpha$ emission line, we masked the data corresponding to the narrow H$\alpha$ + [N II] emission lines. The narrow emission lines were fitted separately after the broad component had been modeled. Therefore, in addition to the components already fitted with Bagpipes, we introduced an extra component to account for the narrow H$\alpha$ + [N II] emission lines.}

{In the Bagpipes code \cite{2018MNRAS.480.4379C}, the AGN continuum is modeled as a broken power law, with the break fixed at a rest-frame wavelength of 5000 $\AA$. The spectral index for wavelengths shortward of the break (\(\lambda_{\text{rest}} < 5000\) \AA) is controlled by a free parameter \(\alpha_{\lambda}\), while another free parameter \(\beta_{\lambda}\) sets the slope longward of the break (\(\lambda_{\text{rest}} > 5000\) \AA). Additionally, a normalization parameter \(f_{5000\text{Å}}\) defines the flux amplitude at \(\lambda_{\text{rest}} = 5000\) \AA. We adopted uniform priors over the ranges \(\alpha_{\lambda} \in [-2, 4]\) and \(\beta_{\lambda} \in [-2, 4]\) for the two spectral indices, and a log-uniform prior over \(f_{5000\text{Å}} \in [10^{-30}, 5\times10^{-20}] \, \rm erg/s/cm^2/\AA\) for the normalization of flux density.
}

{As shown in the bottom panels in the Figs. \ref{fig:8777} and \ref{fig:3651}, we obtained the best-fit model combination incorporating both the stellar continuum and the AGN components. In the figure, the blue solid line represents the observed data, and the black solid line indicates the corresponding observational uncertainties. We show the best-fit model for the broad H$\alpha$ emission line as a red dashed line, the best-fit stellar continuum component as a red dot-dashed line, and the combined model of the stellar component and broad H$\alpha$ emission as a green dotted line. The best-fit narrow emission lines of H$\alpha$ + [N II] are plotted as an orange solid line, while the red solid line represents the total model, which includes contributions from the stellar component, AGN continuum, and narrow emission lines.}

{According to \citet{2005ApJ...630..122G}, the mass of the supermassive black hole can be derived from the broad H$\alpha$ emission line parameters using:}

{
\begin{equation} 
M_{\rm BH} = 2.0 \times 10^6 \, M_{\odot} \times \left( \frac{L_{\rm H\alpha,broad}}{10^{42} \, \rm erg \, s^{-1}} \right)^{0.55} \times \left( \frac{\rm FWHM_{\rm H\alpha,broad}}{10^3 \, \rm km \, s^{-1}} \right)^{2.06},
\label{eq:Mbh_Ha}
\end{equation}
where \( L_{\rm H\alpha,broad} \) is the rest-frame luminosity of the broad H$\alpha$ emission line, which can be obtained by integrating the best-fit model profile converted to the rest frame:
\begin{equation} 
L_{\rm H\alpha,broad} = 4\pi d^2_L \int_{-\infty}^{\infty} n_{\rm broad} \times \frac{1}{\sqrt{2\pi} \, \sigma_{\rm broad}} \times \exp\left[ -\frac{(\lambda - \lambda_{\rm H\alpha,\rm broad})^2}{2\sigma_{\rm broad}^2} \right] d\lambda,
\label{eq:L_Ha} 
\end{equation}
where $d_L$ is the luminosity distance at the redshift of the source.}

{The FWHM\(_{\rm broad}\) of the broad H$\alpha$ line is the full width at half maximum, which can be derived from the width of the best-fit Gaussian model:
\begin{equation} 
{\rm FWHM}_{\rm broad} = 2 \sqrt{2 \sigma_{\rm broad}^2 \ln 2}.
\label{eq:FWHM} 
\end{equation}
}

{Through fitting the broad H$\alpha$ emission line, we obtained best-fit Gaussian line widths of $\sigma_{\rm H\alpha,broad} = 3510.44^{+155.55}_{-163.11}$ km/s and $\sigma_{\rm H\alpha,broad} = 4645.09^{+87.42}_{-100.55}$ km/s for NIRSpec-8777 and ZF-UDS-3651, respectively. These correspond to full width at half maximum (FWHM) values of FWHM $= 8265.42^{+366.29}_{-384.09}$ km/s and FWHM $= 10938.35^{+205.86}_{-236.78}$ km/s. We have listed these values in Table \ref{Tab:fit}. The FWHM of the second source, ZF-UDS-3651, is similar with that of GS-9209—a source also with a quiescent host galaxy—reported as FWHM $= 10300 \pm 700$ km/s in \cite{2023Natur.619..716C}. The similarity between our spectral features and those of GS-9209 suggests that our selected sources are viable AGN candidates in passively evolving galaxies.}

\begin{table}[hbtp]
\centering
\caption{{The emission lines of the two JWST AGNs in this work.} }
\label{Tab:fit}
\begin{tabular}{cccc}
\hline   
\hline   
value & NIRSpec-8777 & ZF-UDS-3651 & units \\ \hline     
 $flux_{H\alpha\rm ,broad}$ & $1.7\pm0.2$ & $4.2\pm 0.4$ & $10^{-19}\,\rm erg/s/cm^2$ \\
 $L_{H\alpha\rm ,broad}$ & $3.9\pm0.2$ & $6.1\pm0.3$ & $10^{40}\,\rm  erg/s$ \\
 FWHM & $8265.42^{+366.29}_{-384.09}$ & $10938.35^{+205.86}_{-236.78}$ & $\rm km/s$ \\
 $flux_{H\alpha \rm ,narrow}$ & $1.8\pm0.1$ & $2.4\pm 0.2$ & $10^{-19}\,\rm erg/s/cm^2$  \\
 $L_{H\alpha\rm ,narrow}$ & $4.1\pm 0.2$ & $3.5\pm 0.2$ & $10^{40}\, \rm erg/s$ \\
 SFR & $0.03\pm 0.01$ & $0.27\pm 0.07$ & $\rm M_{\odot}/yr$ \\
\hline   
\hline   
\end{tabular}
\end{table}

{To estimate the SFR of these galaxies, we followed the method employed in \cite{Lei:2023mke}, using the luminosity of the narrow H$\alpha$ emission line:
\begin{equation}
{\rm SFR}_{{\rm H}\alpha,\rm narrow} = 7.9 \times 10^{-42} \times \left(\frac{L_{{\rm H}\alpha,\rm narrow}}{\rm erg/s}\right)  \; (M_{\odot}/\mathrm{yr}),
\label{eq:SFR}
\end{equation}
where \( L_{{\rm H}\alpha,\rm narrow} \) denotes the luminosity of the narrow H$\alpha$ emission line. The measured flux and derived luminosity of the narrow H$\alpha$ emission line for both sources are summarized in Table \ref{Tab:fit}, with the corresponding SFR values listed in the last column. The low SFR values derived from the narrow H$\alpha$ emission lines of these two sources, coupled with their spectral signatures of an old stellar population, confirm that they belong to the category of passively evolving galaxies. These characteristics make them suitable candidates for testing theories of cosmological coupling between BHs and dark energy.
}

\section{Parameter estimation for the black hole cosmological coupling model}\label{app:b}

As described in \citep{Farrah:2023opk,2011A&A...535A..87S}, several systematic biases can influence on $k$ measurement. To evaluate their impacts, we estimate $k$ in two scenarios, with and without explicit modeling of systematic bias parameters. The systematic bias parameters include

\begin{itemize}
\item $B_{m,M_{\star}}$: the uncertainties of the Initial Mass Function (IMF) in SED modeling;
\item $B_g$: the bias induced by the accretion history;
\item $B_{s,dyn}$: the bias in SMBH mass measurements from stellar dynamics;
\item $B_{s,vir}$: the selection bias toward massive SMBHs;
\item $B_{m,vir}$: the measurement bias of the single-epoch virial mass.
\end{itemize}
Because $B_{m,M_{\star}}$, $B_{m,vir}$, and $B_{s,dyn}$ are primarily determined by observational methodologies and are approximately redshift independent, we adopt the same priors as referred in \citet{Farrah:2023opk}. For $B_{s,vir}$, the prior range adopted in \citet{Farrah:2023opk} remains applicable, even after accounting for the sensitivity of JWST and our stellar-mass cut $M_{\star} > 4\times 10^{10}\, M_{\odot}$. Since $B_g$ may evolve with redshift, \citealt{Farrah:2023opk} estimated a rough range based on SMBH lifetimes at $z \sim 2.7-4.0$ \citep{2021MNRAS.505..649K}. As our sample extends to $z \sim 6$, the corresponding cosmic time increases by $~20\%$. Thus, we consider extra extended prior for $B_{s,vir}$, broadened by $\sim 0.08$ dex.
The detailed priors and posterior constraints are summarized in Table~\ref{Tab:para_k}.

\begin{table}[hbtp]
\centering
\caption{{Priors and posterior results for the parameters in the BH cosmological coupling model.} }

\label{Tab:para_k}
\begin{tabular}{cccc}
\hline   
\hline
\multicolumn{3}{c}{with bias parameter} \\
parameter & prior of parameter inference & posterior \\ \hline     
 $k$ & Uniform(-20,20)  & $-0.16^{+0.39}_{-0.39}$  \\
 $B_{m,M_\star}$ & Uniform(0.08,0.33)   & $0.21^{+0.08}_{-0.09}$ \\
 $B_g$ & Uniform(0,0.1)  &  $0.05^{+0.03}_{-0.03}$ \\
 $B_{s,dyn}$ & Uniform(0,0.3)  & $0.15^{+0.10}_{-0.10}$ \\
 $B_{s ,vir}$ & Uniform(-0.3,-0.1)  & $-0.20^{+0.07}_{-0.07}$ \\
 $B_{m,vir(Mg {\rm\, II})}$ & Uniform(-0.4,0.1)  & $-0.16^{+0.18}_{-0.16}$ \\
 $B_{m,vir(H {\rm\, \beta})}$ & Uniform(-0.2,0.1)  & $-0.06^{+0.10}_{-0.10}$ \\
\hline   
\multicolumn{3}{c}{without bias parameter} \\
\hline   
parameter & prior of parameter inference & posterior \\ \hline     
 $k$ & Uniform(-20,20)  & $-0.48^{+0.27}_{-0.27}$  \\
\hline   
\hline   
\end{tabular}
\end{table}

Without applying any bias corrections, the derived cosmological coupling strength parameter is $k=-0.48 \pm0.27$, which is consistent with \(k = 0\) at the \(2\sigma\) confidence level and deviates from the strong coupling prediction \(k = 3\) by \(\sim  9.3 \sigma\). When systematic bias parameters are included, we obtain $k=-0.16\pm0.39$, which is well consistent with being zero  and excludes the scenario of \(k = 3\) at the \(8.1\sigma\) confidence level. Even adopting the extending prior for $B_g$ ($[0, 0.18]$), we find $k = -0.12 \pm 0.40$, which is statistically indistinguishable from the other results and excludes $k=3$ hypothesis at the $7.7\sigma$ confidence level. These consistent results across different bias treatments demonstrate the robustness of our conclusion.

\section{The $M_{BH}-M_{\star}$ evolutionary tracks of LRDs based on different accretion models}
\label{app:c}

We adopted the following initial conditions of simulation of LRDs based on JWST observations:
\begin{itemize}
     \item Initial stellar masses are distributed as $\log_{10}(M_\star/M_\odot) = 9.05 \pm 0.4$ and black hole masses as $\log_{10}(M_{\rm BH}/M_\odot) = 7.35 \pm 0.5$, matching the observed LRD population \citep{2026ApJ...999...70L,2025NatCo..16.9830T}.
    \item Star formation rates follow a relationship with stellar mass that is consistent with measurements of JWST LRDs: $\log_{10}(\rm{SFR}/(M_{\odot}/yr))=0.7\times \log_{10}(M_{\star}/M_{\odot})-5.3$ with log-normal scatter of 0.8 dex \citep{2026ApJ...999...70L,Speagle:2014loa}.
    \item Black hole accretion rates are assigned using the relation $\log_{10}(\dot{M}_{\rm BH}/(\rm M_{\odot}/yr)) = \log_{10}(M_{\rm BH}/M_{\odot}) -9.0$ with log-normal scatter of 0.5 dex, consistent with observations \citep{2021MNRAS.500..215D,2026ApJ...999...70L}.
\end{itemize}

Besides the constant black hole accretion rate (BHAR) assumption, we discuss three physically motivated accretion models that capture the stochastic and time-dependent nature of SMBH growth in realistic galaxy evolution: (1) Episodic Accretion, (2) Exponentially Declining Accretion, (3) Merger-triggered Bursts and (4) Super-Eddington Accretion.

We use Monte Carlo simulations to track the time evolution of individual BHs under these different accretion histories. The mathematical formulations and physical motivations of the models are summarized below.

\begin{figure}[htbp!]
    \centering
    \includegraphics[width=0.475\textwidth]{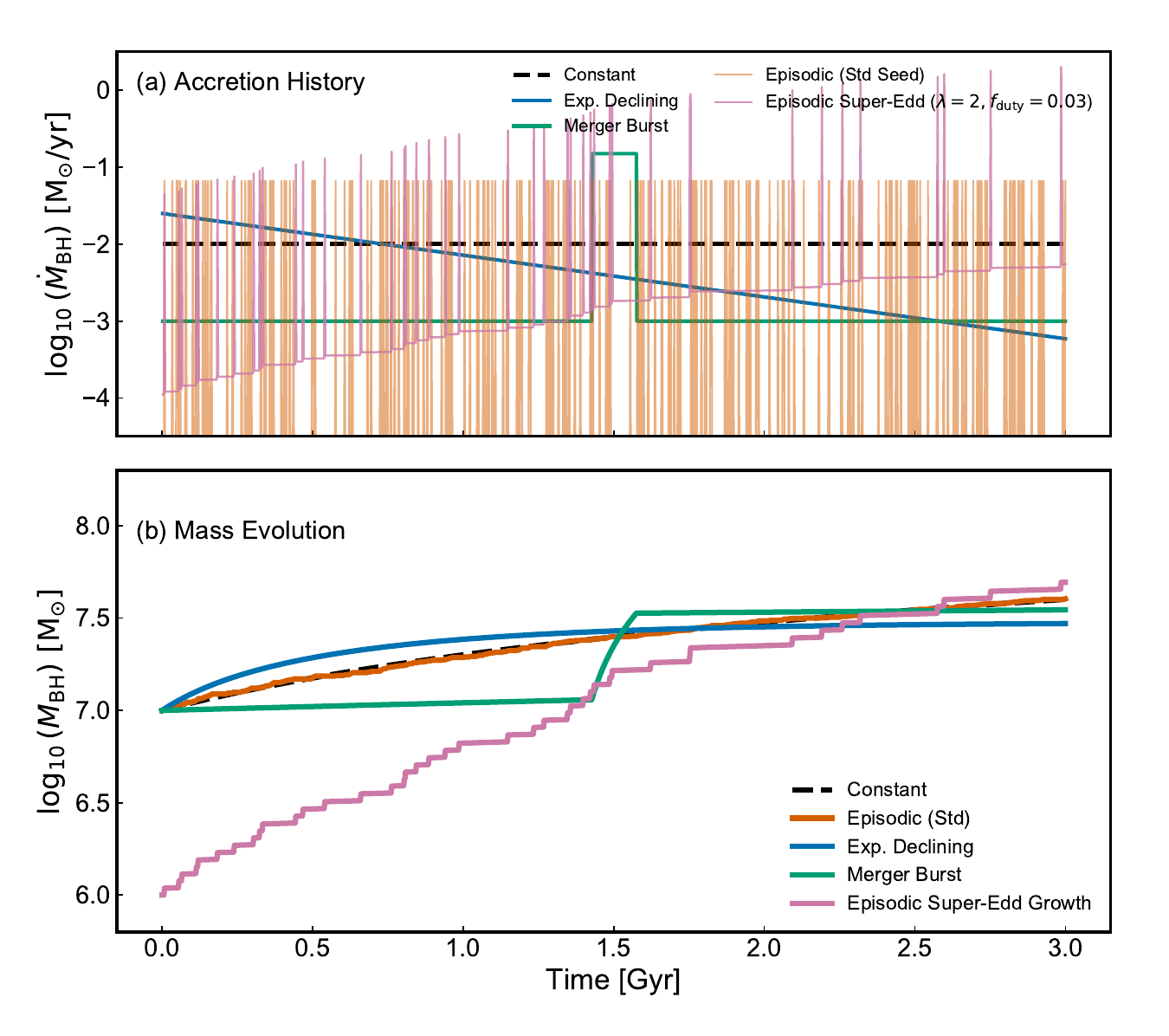}
    \caption{Schematic illustration of the time evolution of Black Hole Accretion Rate (top) and Black Hole Mass (bottom) for the tested models: Constant (black), Episodic (orange), Exponentially Declining (blue), and Merger-triggered (green).}
    \label{fig:R1}
\end{figure}

\begin{itemize}
    \item Model A: Episodic accretion \\
    Motivated by the ANG variability studies (e.g., \citet{2004ApJ...613..109H,2014ApJ...782....9H}), we assume that BH accretion proceeds stochastically. 
    The time-averaged growth is presented by
    \begin{equation}
        \Delta M_{\rm BH} \approx \sum_i \dot{M}_{\rm active} \cdot \delta_i \cdot \Delta t,
    \end{equation}
    where $\delta$ denotes the duty cycle of the $i$-th active episode drawn from a random distribution and $\Delta t$ is the corresponding duration.
    The instantaneous accretion rate during active phases is taken to be relatively high ($\log \dot{M}_{\rm active} \sim \log M_{\rm BH} - 8.3$), such that the integrated mass growth matches of the constant accretion model.

    \item Model B: Exponentially declining \\
    To simulate the feedback-regulated ``Declining'' phase (e.g., \citet{2005Natur.433..604D,2002MNRAS.335..965Y}), we model the accretion rate as an exponential decay following an initial activity peak,
    \begin{equation}
        \dot{M}_{\rm BH}(t) = \dot{M}_{\rm peak} \cdot \exp\left(-\frac{t}{\tau}\right),
    \end{equation}
    where $\tau$ is the e-folding timescale, taken to lie in the range of $100-500$ Myr. 
    This model can effectively halt the black hole growth at later times ($z \lesssim 2$).

    \item Model C: Merger-triggered bursts \\
    Following the light-curve models from hydrodynamic simulations (e.g., \citet{2006ApJS..163....1H}), we implemented a scenario where the bulk of the mass growth occurs during short, intense accretion episodes triggered by galaxy mergers, followed by quiescent phases. The accretion rate is modeled as
    \begin{equation}
        \dot{M}_{\rm BH}(t) = 
        \begin{cases} 
        \dot{M}_{\rm burst}, & \text{if } |t - t_{\rm merger}| < t_{\rm burst}, \\ 
        \dot{M}_{\rm quiescent}, & \text{otherwise},
        \end{cases}
    \end{equation}
    where $t_{\rm burst} \sim 100$ Myr represents the visible quasar lifetime.

\begin{figure*}[htbp]
    \centering
    \includegraphics[width=0.4\textwidth]{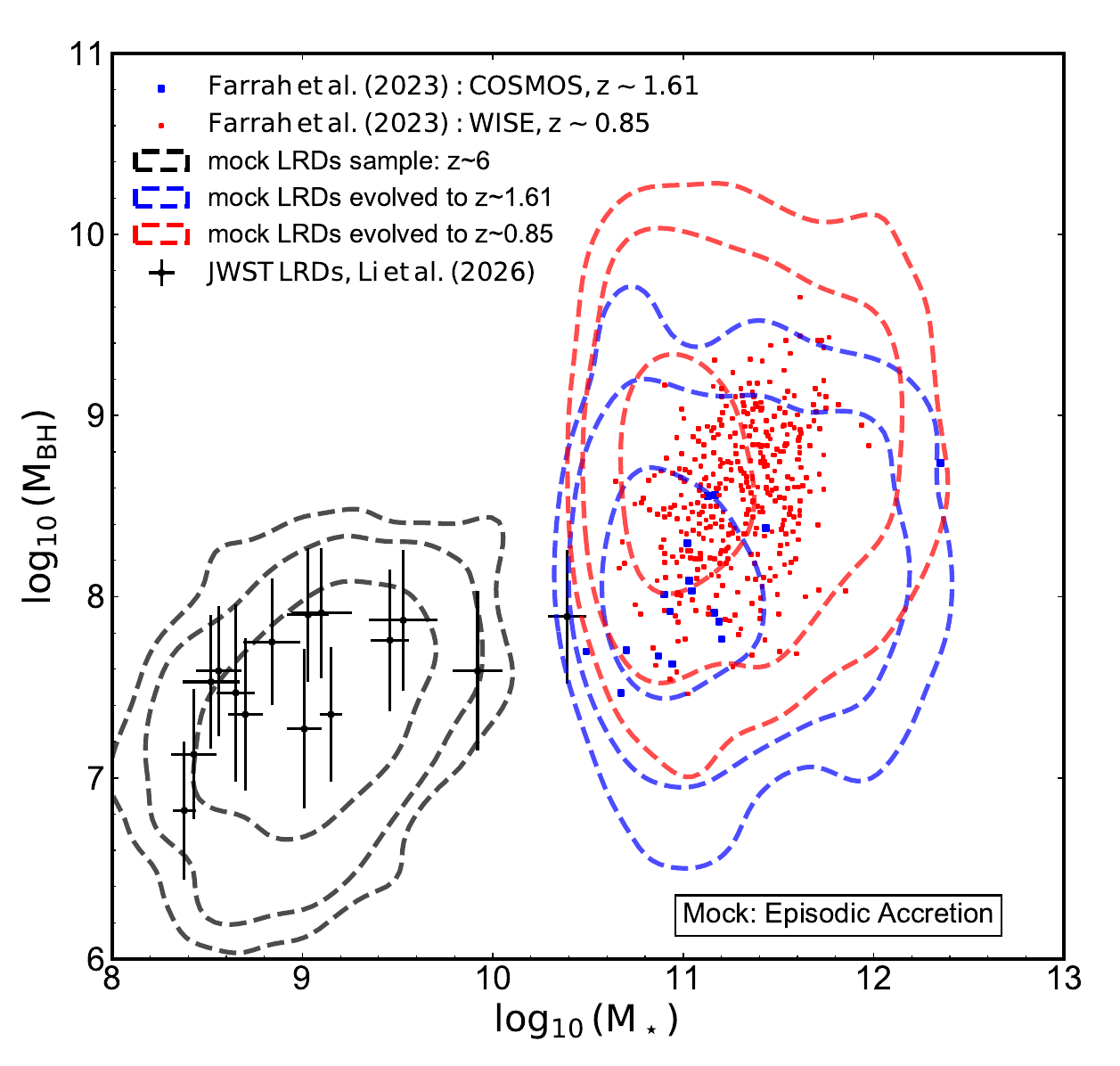}
    \includegraphics[width=0.4\textwidth]{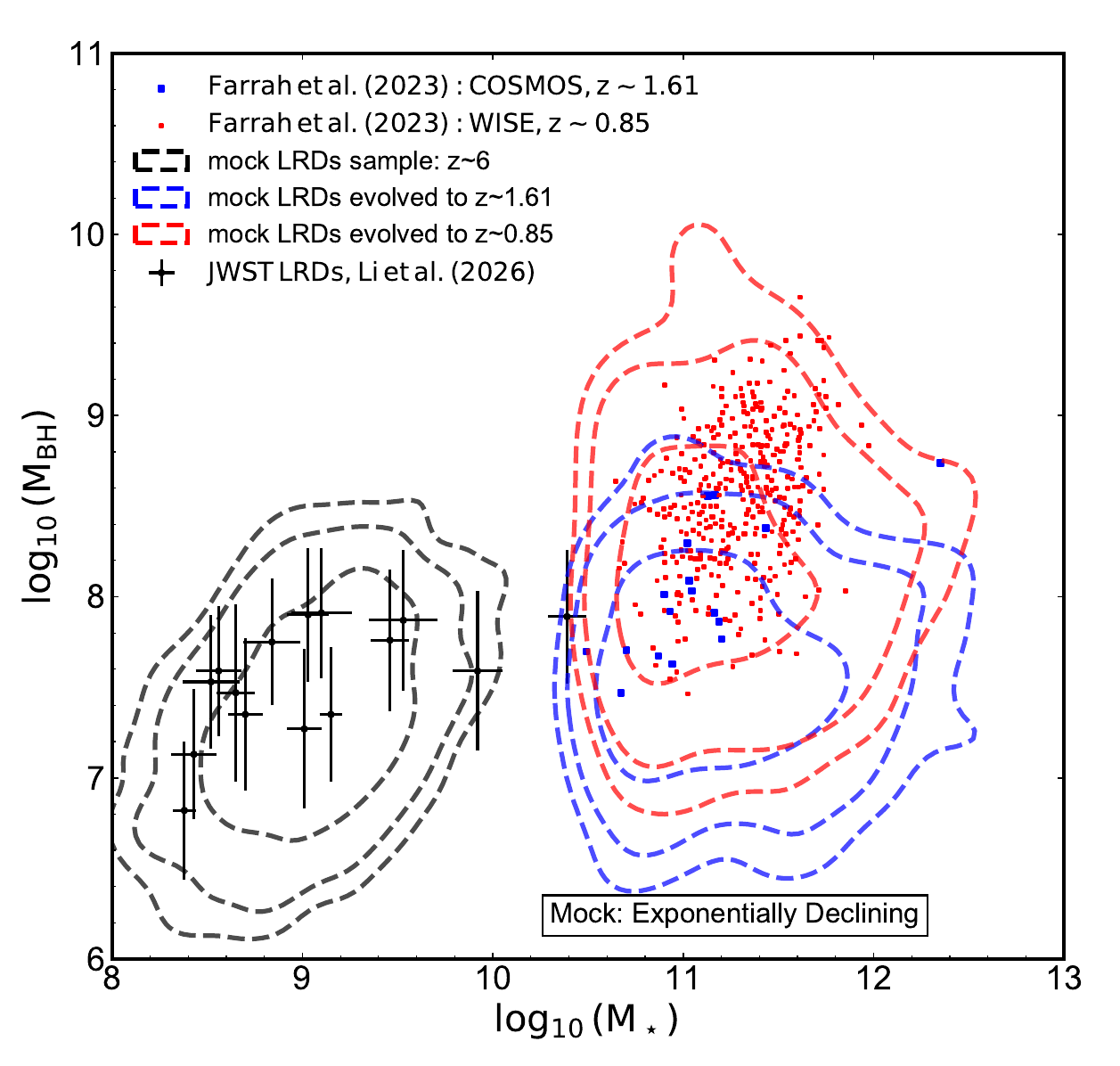}
    \includegraphics[width=0.4\textwidth]{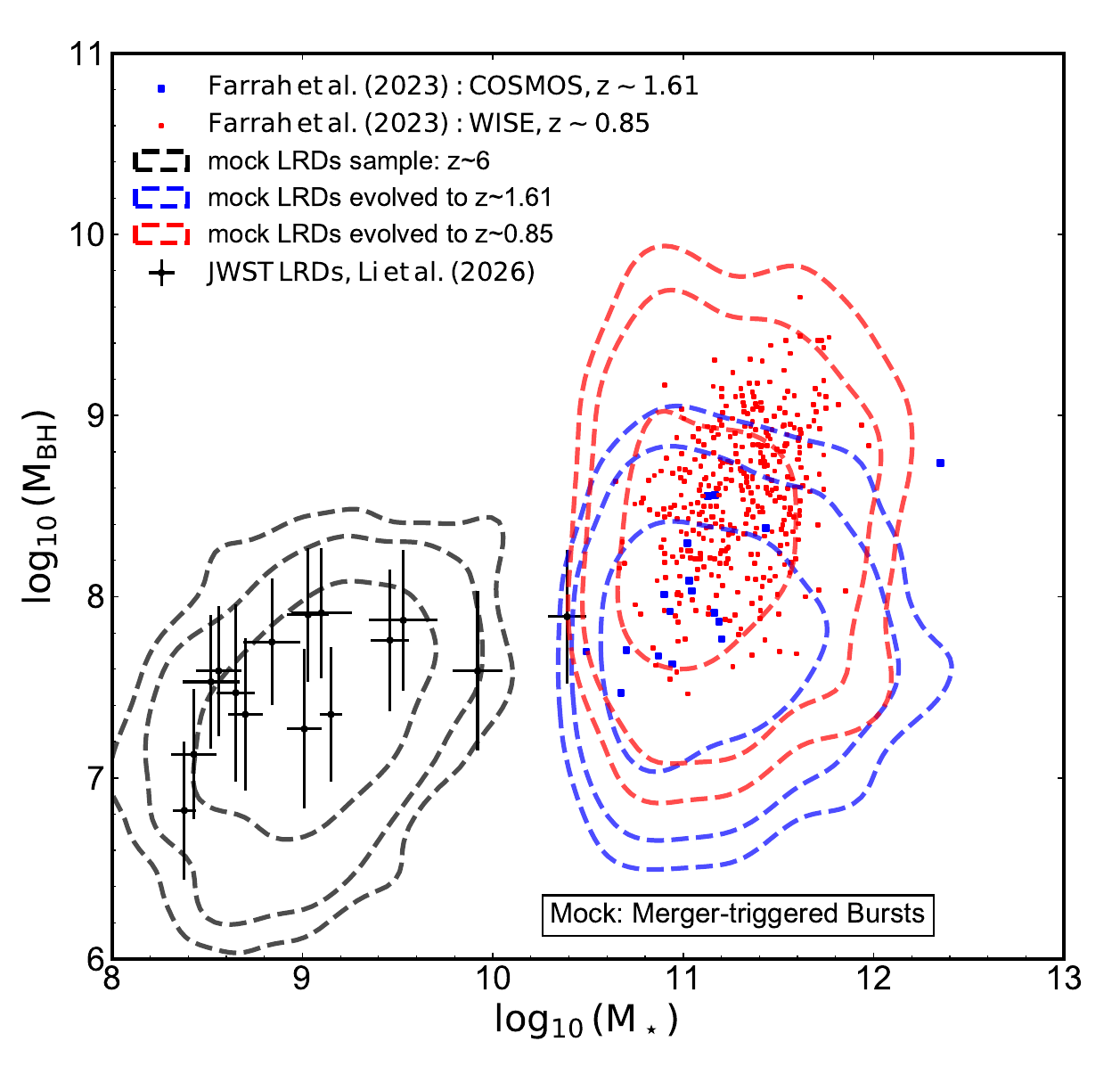}
    \includegraphics[width=0.4\textwidth]{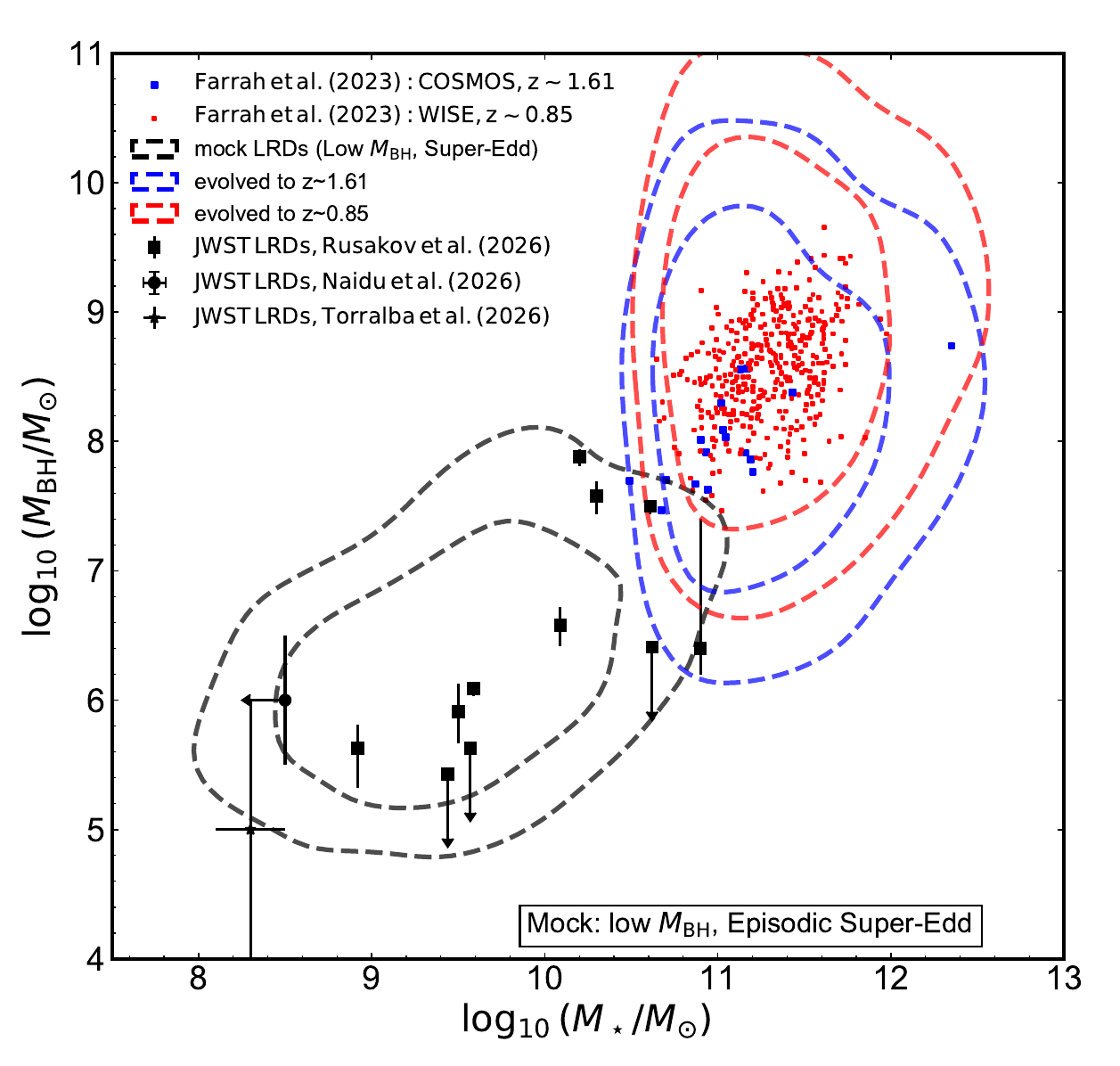}
    
    \caption{The evolved $M_*-M_{BH}$ plane under different accretion assumptions. The contours represent the density distribution of the simulated population. The black data points are taken from \cite{2026ApJ...999...70L,2026arXiv260630711N,2026ApJ..1005L..37T,2026Natur.649..574R}.}
    \label{fig:R2}
\end{figure*}

    \item Model D: Super-Eddington accretion\\
    LRDs may host intermediate-mass BHs accreting at high or super-Eddington rates (e.g., \cite{2026arXiv260630711N,2026arXiv260317667M,2026ApJ..1005L..37T,2025arXiv250316596N,2026Natur.649..574R,2026ApJ...999..183U,2026ApJ...996..129G,2025ApJ...980L..34L,2026ApJ..1005..150L}), which naturally explains their observed X-ray weakness \citep{2026arXiv260222386M,2025PASJ...77..811I,2025ApJ...988L..22I,2024ApJ...976...96P}.
    In our Super-Edd test, we used an accretion rate 1.5 dex higher, and the average black hole mass of the samples is an order of magnitude lower than the original overall mass. The mass growth driven by intermittent super-Eddington bursts is presented by
    \begin{equation}
    \Delta M_{\rm BH} \approx \sum_i \left( \lambda \frac{M_{\rm BH}}{\rm 45\, Myr} \right) \cdot \delta_i \cdot \Delta t,
    \end{equation}
    where $\lambda=2.0$ is the super-Eddington ratio. To model the transition of the accretion mode across cosmic time, we assume the instantaneous accretion rate during active phases drops from a super-Eddington regime ($\lambda \approx 2.0$) at high redshift ($z \sim 6$) to a sub-Eddington regime ($\lambda \approx 0.7$) at lower redshifts ($z \sim 1.6$).
\end{itemize}

Figure~\ref{fig:R1} illustrates the individual growth tracks of a seed black hole under these different scenarios over the simulated timescale (3 Gyr). We then apply these prescriptions to the full LRD sample. Figure~\ref{fig:R2} shows the resulting $M_*-M_{BH}$ distribution at $z \sim 1.6$ and $z \sim 0.85$ under these relaxed accretion assumptions.
Despite the different mass-assembly pathways (continuous vs. bursty),
the final locus of the population largely remains within the parameter space of $z\sim2$ ETGs \citep{Farrah:2023opk} and local massive galaxies.
Therefore, we conclude that our primary conclusion, LRDs are plausible progenitors of ETGs, remains robust against variations in the detailed SMBH accretion history.

\clearpage

\end{appendix}
\end{CJK*}

\end{document}